\documentclass[floats,floatfix,showpacs,amssymb,prd,twocolumn,superscriptaddress,nofootinbib,nolongbibliography,reprint,preprintnumbers]{revtex4-1}

\usepackage{amssymb,amsmath,verbatim,mathtools,needspace,enumitem,etoolbox,graphicx,physics,microtype,afterpage,xspace,tabularx,lmodern,multirow,braket}
\usepackage{gensymb}
\usepackage[normalem]{ulem}
\usepackage[dvipsnames, usenames]{xcolor}
\definecolor{linkcolor}{rgb}{0.0,0.3,0.5}
\usepackage[unicode, colorlinks=true, linkcolor=linkcolor, citecolor=linkcolor, filecolor=linkcolor, urlcolor=linkcolor, linktocpage, breaklinks]{hyperref}
\usepackage[all]{hypcap}
\usepackage{subfigure}
\usepackage[T1]{fontenc}
\usepackage[utf8]{inputenc}
\usepackage[usenames,dvipsnames]{xcolor}
\hypersetup{colorlinks=true,citecolor=romared,linkcolor=romared,urlcolor=romared}

\definecolor{romared}{RGB}{142,0,28}

\newcommand{\be}{\begin{equation}}
\newcommand{\ee}{\end{equation}}

\def\be{\begin{equation}}
\def\ee{\end{equation}}
\newcommand{\beq}{\begin{eqnarray}}
\newcommand{\eeq}{\end{eqnarray}}

\usepackage{aas_macros}
\usepackage{makecell}
\usepackage{soul}
\usepackage{amssymb}
\usepackage{longtable}

\usepackage{lipsum}

\newcolumntype{Y}{>{\centering\arraybackslash}X}

\begin{document}
\title{Reconstruction of ringdown with excitation factors}
\author{Naritaka Oshita}
\affiliation{Center for Gravitational Physics and Quantum Information, Yukawa Institute for Theoretical Physics, Kyoto University, 606-8502, Kyoto, Japan}
\affiliation{The Hakubi Center for Advanced Research, Kyoto University,
Yoshida Ushinomiyacho, Sakyo-ku, Kyoto 606-8501, Japan}
\affiliation{RIKEN iTHEMS, Wako, Saitama, 351-0198, Japan}
\author{Vitor Cardoso} 
\affiliation{Niels Bohr International Academy, Niels Bohr Institute, Blegdamsvej 17, 2100 Copenhagen, Denmark}
\affiliation{CENTRA, Departamento de F\'{\i}sica, Instituto Superior T\'ecnico -- IST, Universidade de Lisboa -- UL, Avenida Rovisco Pais 1, 1049-001 Lisboa, Portugal}

\preprint{YITP-24-81, RIKEN-iTHEMS-Report-24}

\begin{abstract}
In black hole perturbation formalism, the gravitational waveform is obtained by the convolution of the Green's function and the source term causing radiation emission. Hence, the ringdown properties, namely its start time, depend on both functions. The unknown time-shift encoded in the Green's function introduces a ``time-shift problem'' for ringdown. We study the ringdown time-shift problem by reconstructing a waveform via the excitation factors of quasi-normal modes (QNMs) of a spinning black hole. For the first time, we reconstruct ringdown with a significant number of QNMs weighted with their excitation factors and confirm its excellent convergence. 
We then precisely identify the ringdown starting time. We also find (i) that for moderate or large spins and $\ell=m=2$, QNMs should be included up to around the $20$th prograde overtones and around fifth retrograde overtones to reconstruct the ringdown waveform for the delta-function source with a mismatch threshold ${\cal M} < {\cal O}(10^{-3})$. For higher angular modes, a more significant number of QNMs are necessary to reconstruct it; (ii) that the time shift of ringdown caused by the Green's function is the same for different $(\ell, m, n)$ modes but that nontrivial sources can change this conclusion. Finally, we demonstrate (iii) that the greybody factor can be reconstructed with the superposed QNM spectrum in the frequency domain.
\end{abstract}

\maketitle

\section{Introduction}
A perturbed black hole (BH) -- such as the end-product of a compact binary merger -- relaxes to its final quiescent state by emitting gravitational radiation in a set of characteristic frequencies. This relaxation stage is often called the ``ringdown,'' when the wave amplitude decays exponentially with time. The ringdown waveform is expressed by a superposition of multiple BH quasinormal modes (QNMs)~\cite{Vishveshwara:1970zz,Chandrasekhar:1975zza,Berti:2009kk}. Because the signal fades exponentially, it is important to understand its start time to maximize the signal-to-noise ratio of multiple QNMs, and to use it to perform tests of fundamental issues.
The determination of ringdown start time is plagued with ambiguities and subtleties, some caused by nonlinear effects during BH excitation~\cite{Baibhav:2023clw}. In the linear regime, given a source term and a full gravitational waveform, the ringdown start time can in principle be determined (see more below), with a sensible {\it definition} and provided power-law tails have a negligible impact~\cite{Baibhav:2023clw}. 
However, identifying the start time of ringdown is still an open problem as there is no explicit way to read it from the Green's function of BH perturbations, an issue related to the time-shift problem of ringdown~\cite{Andersson:1996cm,Nollert:1998ys,Sun:1988tz,Berti:2006wq}. 
As was discussed in Ref.~\cite{Andersson:1996cm}, the time-shift and ringdown starting time problems arise when one tries to associate initial data with the excitation coefficient of each QNM. This is possible only after the relevant feature in the initial data has scattered at the potential barrier around the BH. Before it reaches the barrier, we would not expect to see the excitation of QNMs.
However, the location at which QNMs are excited is fuzzy, as the potential barrier has a finite width and its shape depends on the multipole modes. As such, it is non-trivial to determine exactly when QNMs are excited. Also, as each QNM amplitude is exponentially larger at earlier times, this problem is relevant to the convergence of the QNM expansion of ringdown as well \cite{Leaver:1986gd,Andersson:1996cm}.

Here, we attack this problem by utilizing the excitation factors~\cite{Leaver:1985ax,Leaver:1986vnb,Leaver:1986gd,Sun:1988tz,Nollert:1998ys,Andersson:1995zk,Berti:2006wq,Zhang:2013ksa,Oshita:2021iyn}, which are the residue of a pole (the QNM) of the Green's function, and quantify the excitability of each QNM. We reconstruct the time-domain waveform with the excitation factors and QNMs and then identify the start time of the ringdown.
Reconstruction of the waveform at early time was first performed in a Schwarzschild background~\cite{Leaver:1986gd} (see also Ref.~\cite{Andersson:1996cm}) and was later extended to the Kerr case~\cite{Berti:2006wq} with several QNMs. We will show that it is necessary to include a significant number of QNMs to reconstruct the whole ringdown waveform of a Kerr BH in the mismatch threshold ${\cal M} < {\cal O}(10^{-3})$, e.g., up to around the 20th prograde overtone and up to around the fifth retrograde overtone for $j=0.7$.\footnote{The inclusion of retrograde modes in the QNM fitting has been discussed and performed in e.g. Ref. \cite{Zhu:2023mzv}.}
To our knowledge, we are the first to reconstruct the waveform of ringdown, including the earliest part, by utilizing multiple QNMs for spinning BHs and by confirming the convergence of the reconstructed waveform against the number of included prograde and retrograde QNMs. We then identify the start time of ringdown in the linear regime precisely.
We also for the first time show that for non-trivial source terms, the start time of ringdown may be different among different angular modes and overtones.

Our paper is organized as follows. 
In Sec. \ref{sec_ef}, we briefly review the excitation factors and the time-shift and the starting time problems of ringdown.
In Sec. \ref{sec_fix_time}, we identify the start time of ringdown by assuming the delta-function source term for various values of the spin parameter. We will also show that many QNMs, both prograde and retrograde modes, are needed to reconstruct the waveform precisely.
Finally, we conclude that the time shift caused by the Green's function is the same for different multipole modes and overtones.
In Sec. \ref{sec_angular_dependence}, we will demonstrate and discuss that for non-trivial source terms, the start time of ringdown may depend on the angular modes and even overtone numbers.
In Sec. \ref{sec_gray}, we reconstruct the ringdown spectral amplitude, which has the exponential decay at higher frequencies, with the superposed QNM spectrum including a significant number of prograde and retrograde QNMs.
In Sec. \ref{sec_conclusion}, we provide our conclusion.
Throughout the paper, we set the BH mass to $M=1/2$ and take unit $c=1=G$.

\section{Excitation factors and the time shift problem}
\label{sec_ef}

The response of a BH spacetime to an external perturbation can be expanded around the stationary background. At linearized level, all the information about the dynamical response is encoded in the Teukolsky wavefunction. The waveform at any point is a convolution between the source and the retarded Green's function, and one can identify three main contributions~\cite{Leaver:1986gd,Berti:2006wq}. Here we focus on the ringdown stage.

Consider a perturbation $\tilde{X} (\omega)$ governed by the equation
\be
\frac{d^2\tilde{X}}{dr_*^2}+V\tilde{X}=I\,,\label{eq:SN_like}
\ee
where $r_*$ is the tortoise coordinate, $V$ is a potential barrier, and $I$ is a source term.
We defined the Laplace transform of any quantity as
\be
X(u)=\frac{1}{2\pi}\int_{-\infty+ic}^{+\infty+ic}\tilde{X}e^{-i\omega t} d\omega\,,
\ee
with $u = t-r_*$ and the source term 
\be
I=i\omega X(t=0)-\left. \frac{\partial X}{\partial t} \right|_{t=0} \,.\label{eq:source_wave}
\ee
Far away from the source, where detectors are located,
\beq
\begin{split}
\tilde{X}&=\int_{-\infty}^{+\infty} dr_*' I G (\omega, r_*, r_*')\\
&=\frac{e^{i\omega r_*} A_{\rm out}}{2i\omega A_{\rm in}}\int_{-\infty}^{+\infty} dr_*' \frac{I \tilde{X}_{r_+} (\omega, r_*')}{A_{\rm out}}\,.
\end{split}
\eeq
where $G$ is the Green's function
\be
G(\omega, r_*, r_*') = \frac{e^{i\omega r_*}}{2i\omega A_{\rm in}} \tilde{X}_{r_+} (\omega, r_*'),
\ee
and $\tilde{X}_{r_+}$ is the homogeneous solution to \eqref{eq:SN_like}, with the asymptotic behavior,
\beq
\tilde{X}_{r_+}=
\begin{cases}
&e^{-i (\omega-m\Omega)r_{\ast}}\,,\qquad r\to r_+\\
&A_{\rm in}e^{-i\omega r_*}+A_{\rm out}e^{i\omega r_*}\,,\quad r_*\to \infty.
\end{cases}
\label{homo_sol_Ain_Aout}
\eeq
Note that the tortoise coordinate $r_*$ is defined up to an integration constant $c_*$, $r_* \to r_* +c_*$, which leads to an ambiguity in the phase of $A_{{\rm in}/{\rm out}}$\footnote{To be clear, since this one of the main points of this work, the ambiguity is unimportant if source and excitation factors are calculated with the same definition of tortoise coordinate; then, it is equivalent to a re-parametrization of time, and the problem is completely defined by the initial data. However, it is often the case that the excitation factors were calculated separately and using different conventions for the tortoise coordinate.}.
Defining the following two functions:
\beq
\tilde{X}_{\rm G}(\omega) &=& \frac{e^{i\omega r_*} A_{\rm out}}{2i\omega A_{\rm in}}\,,
\label{green_waveform}\\
\tilde{X}_{\rm T} (\omega) &=& \int_{-\infty}^{+\infty} dr_*' \frac{I (\omega, r_*') \tilde{X}_{r_+} (\omega, r_*')}{A_{\rm out}}\,,
\label{source_waveform}
\eeq
we obtain the time domain waveform $X(u)$ by taking the convolution of
\be
X(u) = X_{\rm G}(u) * X_{\rm T}(t)\,,
\ee
where $X_{\rm G}(u)$ and $X_{\rm T}(t)$ are the inverse Laplace transforms of $\tilde{X}_{\rm G} (\omega)$ and $\tilde{X}_{\rm T} (\omega)$, respectively.
The QNM signature is imprinted in $X_{\rm G}$ as the poles in (\ref{green_waveform}), i.e., zeros of $A_{\rm in} (\omega)$, lead to the QNM excitation. The waveform $X_{\rm G}$ at late times can be expressed as
\begin{align}
\begin{split}
X_{\rm G} (u) &\equiv
\int_{-\infty}^{\infty} d \omega \frac{e^{-i\omega u} A_{\rm out}}{2i\omega A_{\rm in}}\\
&= \sum_n E_ne^{-i\omega_n (u-t_*)} \equiv X_{\rm E} (u)\,\, \text{for}\,\, u \geq t_{\ast}\,,
\end{split}
\label{sum_qnm}
\end{align}
where $n$ is the overtone number, the excitation factor $E_n$ is 
\beq
E_n=\frac{A_{\rm out}(\omega_{n})}{2\omega\alpha_n} e^{2i \omega_n c_*}\,,
\eeq
with $\alpha_n \equiv (dA_{\rm in} / d\omega)_{\omega = \omega_{n}}$ and $c_*$ is the ``ambiguity parameter'' to be determined so that $E_n$ gives the initial amplitude of the QNM in $X_{\rm G}$. Indeed, as noted previously, the ratio of the asymptotic amplitudes in (\ref{homo_sol_Ain_Aout}), $A_{\rm out}/A_{\rm in}$, has an ambiguity in its phase due to an integration constant that comes in the definition of $r_*$.
Therefore, the excitation factor $E_n$ has an ambiguity in its phase and amplitude quantified by $e^{2i \omega_n c_*}$.
Correspondingly, an unknown constant $t_*$ is encoded in the Green's function or in $X_{\rm G}$ and contributes to the determination of the start time of QNM excitation.

Hereinafter, we refer to the ringdown in $X_{\rm G} (u)$ as the {\it fundamental ringdown} in which the amplitude of each QNM is determined by the excitation factors.
Finding the value of $t_*$ in the fundamental ringdown is equivalent to properly fixing the ambiguous factor $e^{2i \omega_n c_*}$ in the excitation factors as the factor of $e^{i \omega_n (2c_*)}$ leads to a time shift.
\begin{figure}[t]
\centering
\includegraphics[width=1\linewidth]{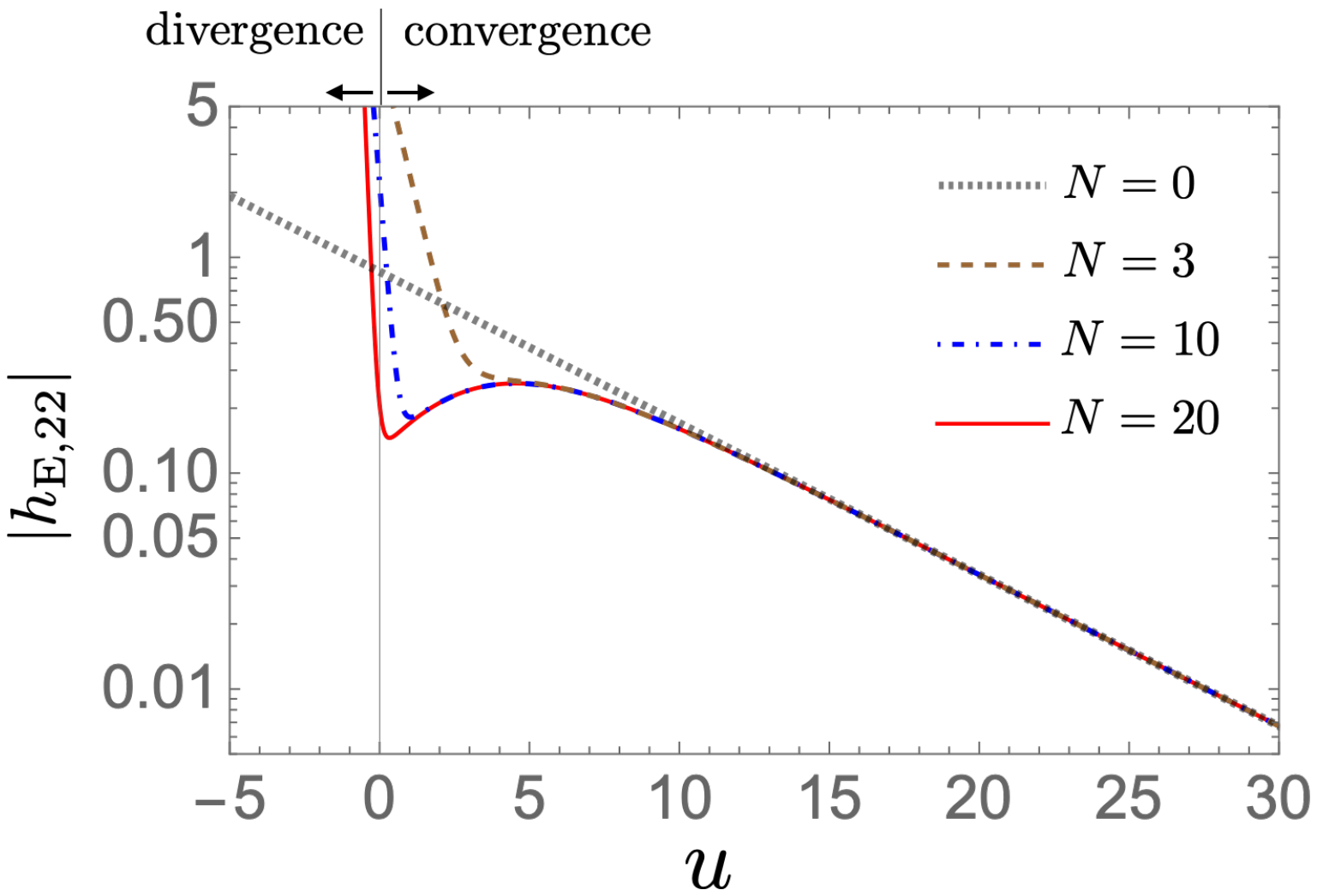}
\caption{The demonstration of the late-time convergence and early-time divergence of $h_{{\rm E},\ell m}$ for a Kerr BH with the spin parameter of $j=0.7$ and $\ell=m=2$. The Kerr prograde QNMs are included up to the $N$-th tone, and the excitation factors obtained in Ref. \cite{Oshita:2021iyn} are used.}
\label{pic_divcon}
\end{figure}

An actual ringdown waveform is obtained by taking the convolution of $X_{\rm G}$ and $X_{\rm T}$. The function $X_{\rm T}$ deforms the fundamental ringdown by changing the amplitude and phase of each QNM in $X_{\rm G}$. As a simple example, we take a source term $I$ localized far from the BH $x_s (\gg r_+)$ with an amplitude
\be
I = \delta (r_* - x_s)\,,
\label{delta_source_term}
\ee
we have
\be
\tilde{X}_{T} (\omega_{\ell mn}) = e^{i \omega_{\ell mn} x_s}\,,
\ee
and the ringdown part of the full waveform $X (u)$ reduces to
\beq
X(u) = \sum_{n} E_n e^{-i\omega_n (u -x_s - t_*)} = X_{\rm E}(u-x_s) \,.
\label{excitation_model}
\eeq
This result is sensible, as the source located at $r_* = x_s$ takes a time $t\sim x_s$ to reach the light ring at $r_* \sim 0$ and excite the QNMs. However, the precise location at which QNMs are excited is fuzzy, and the exact start time of QNM excitation is unclear. Such fuzziness is part of the origin of the ambiguity of $t_*$ in (\ref{sum_qnm}), which is independent of a source term we take but is encoded in the Green's function.
Also, the superposed QNM model (\ref{sum_qnm}) at $u< t_*$ or (\ref{excitation_model}) at $u<x_s+ t_*$ may diverge as overtones' amplitudes diverge exponentially at earlier times~\cite{Leaver:1986gd,Andersson:1996cm,Berti:2006wq}. 
Thus, the faithful representation of a signal via a QNM expansion is a non-trivial problem. In particular, what is the time after (before) which the QNM expansion converges (diverges), as illustrated Fig. \ref{pic_divcon}?

We here study this problem by reconstructing the fundamental ringdown in $X_{\rm G}(u)$ with the excitation factors and QNMs. We then search for the value of $t_*$ such that
\be
X_{\rm G} (u) = X_{\rm E}(u) \,,\quad \text{for} \quad u>t_*\,,
\ee
within the mismatch threshold ${\cal M} \lesssim {\cal O}(10^{-3})$.
We will show that the $t_*$ is indeed encoded in the Green's function, i.e. one can solve the starting time problem of ringdown as well, and that a significant number of QNMs of the prograde and retrograde modes are necessary to reconstruct the fundamental ringdown for a Kerr BH.

\section{fixing the time shift of the excitation factors}
\label{sec_fix_time}

Let us consider the Kerr perturbation governed by the Teukolsky equation, whose in-mode homogeneous solution ${}_{-2} R_{\ell m}^{\rm (in)}$ satisfies
\be
{}_{-2}R_{\ell m}^{\rm (in)} =
\begin{cases}
&\Delta^2 e^{-i k_{\rm H} r_*}\,, \ (r_* \to -\infty)\\
&r^3 A^{\rm (T)}_{{\rm out},\ell m} e^{i\omega r_*} + r^{-1} A^{\rm (T)}_{{\rm in},\ell m} e^{-i\omega r_*}\,, \ (r_* \to \infty)
\end{cases}
\ee
where $\Delta \equiv r^2 -r + a^2$, $k_{\rm H} \equiv \omega - m \Omega_{\rm H}$, $a=j/2$ is the spin parameter, and $\Omega_{\rm H}$ is the horizon velocity. 
We here also assume the conventional definition of $r_*$ for the Kerr background
\be
r_* \equiv r + \frac{r_+}{r_+ -r_-} \log \left( r-r_+ \right) -\frac{r_-}{r_+ -r_-} \log \left( r-r_- \right)\,,
\label{conventional_tortoise}
\ee
where $r_{\pm}$ are the roots of $\Delta(r) =0$ and $r_+ \geq r_-$.
\begin{figure}[t]
\centering
\includegraphics[width=1\linewidth]{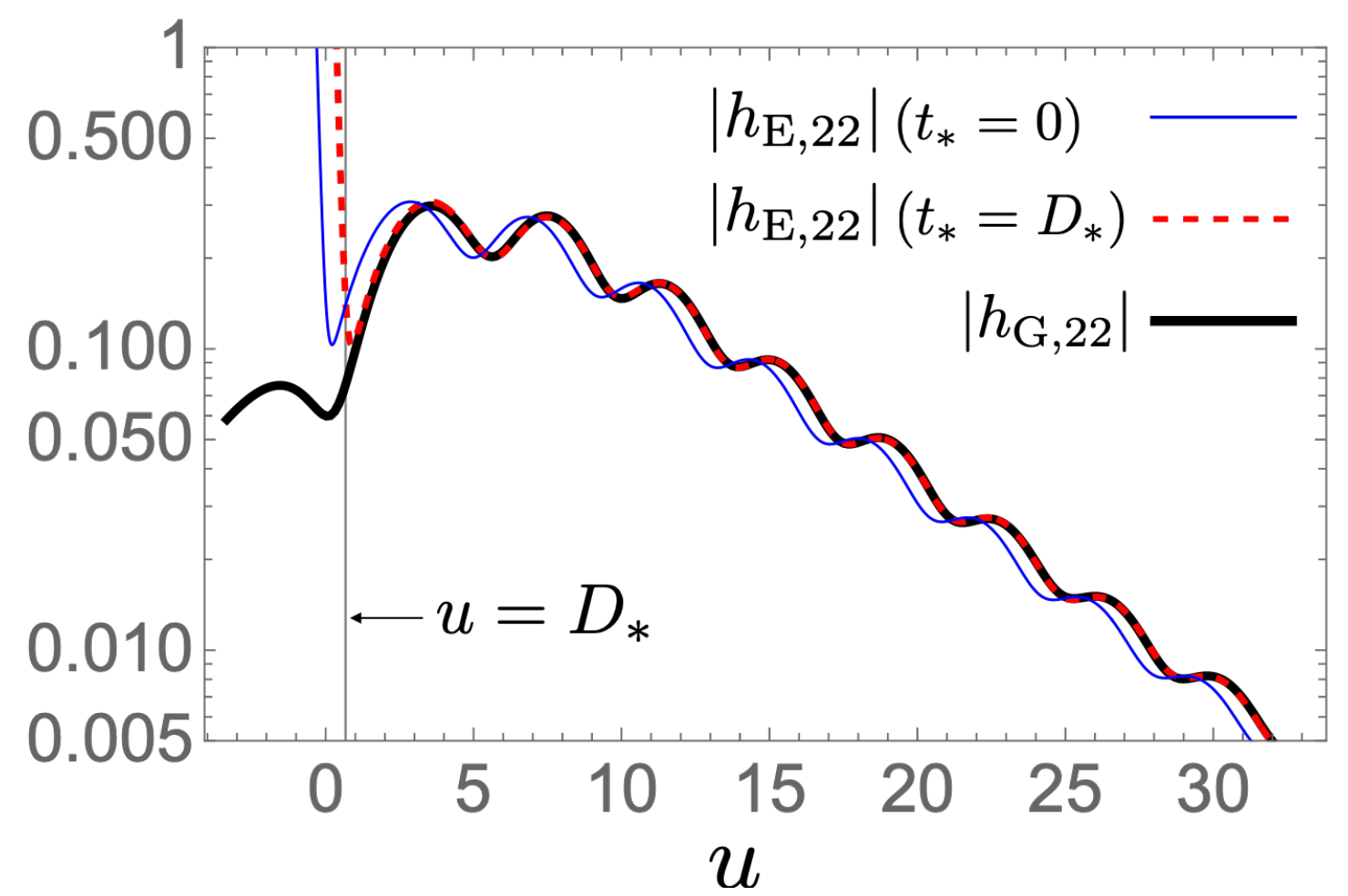}
\caption{Comparison between $h_{{\rm G},22}$ computed with the conventional tortoise coordinate $r_*$ (black thick line) and $h_{{\rm E},22}$ with $t_* = D_*$ (red dashed). Blue thin solid is $h_{{\rm E},22}(u)$ with $t_*=0$. We here set the spin parameter as $j=0.7$. The mismatch is ${\cal M} \simeq 6 \times 10^{-4}$ with $t_{\rm ini} = D_* = t_*$ and $t_{\rm end}=50$. Formally, the source interacts with the light ring at $u\sim 0$; the ringdown starting time we find is consistent with the expectation that it is excited at the light ring (see also Footnote {\ref{footnote_d_star}}).}
\label{pic_time_shift}
\end{figure}
\begin{figure}[t]
\centering
\includegraphics[width=1\linewidth]{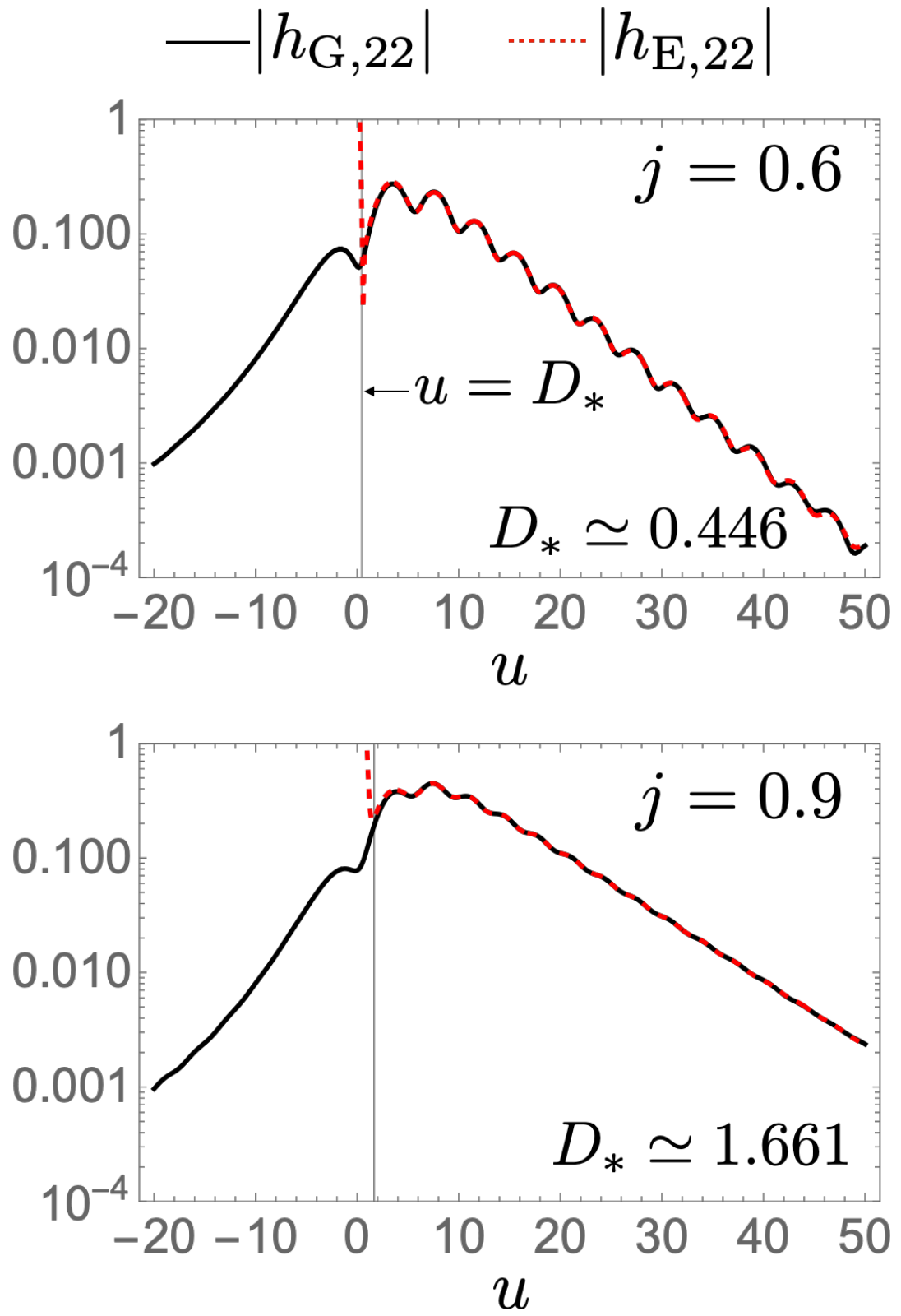}
\caption{Comparison between $h_{{\rm G},22}$ (black) and $h_{{\rm E},22}$ with $t_* = D_*$. We include modes with $(N_{\rm P}, N_{\rm R})=(20,6)$ and $(20,3)$ for $j=0.6$ and $j=0.9$, respectively. The mismatch is ${\cal M} \simeq 3 \times 10^{-3}$ and ${\cal M} \simeq 4 \times 10^{-4}$ for $j=0.6$ and $j=0.9$, respectively, with $t_{\rm ini} = D_*$ and $t_{\rm end}=50$. Our results for the time shift $D_*$ are consistent with Eq.~\eqref{D_ana}, and therefore with the different tortoise coordinate used in the evaluation of the excitation factors in Ref.~\cite{Oshita:2021iyn}.}
\label{pic_starttime}
\end{figure}
\begin{figure}[t]
\centering
\includegraphics[width=1\linewidth]{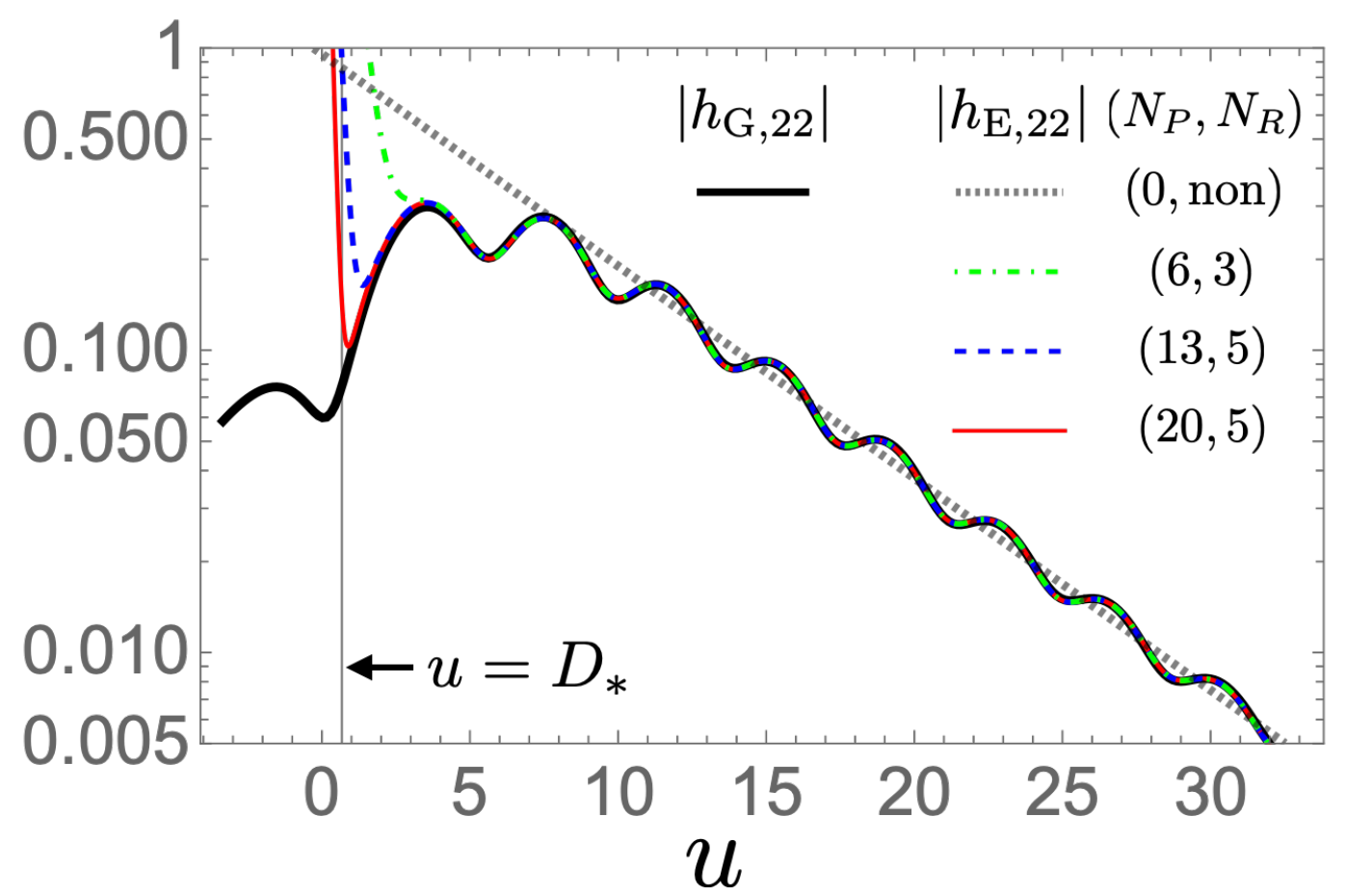}
\caption{Comparison between $h_{{\rm G},22}(u)$ (black) and $h_{{\rm E},22}(u)$ for various values of $N_{\rm P}$ and $N_{\rm R}$.  $N_{\rm R}= \text{non}$ means that the retrograde modes are not included in $h_{{\rm E},\ell m}$. The vertical lines indicate the value of $u = D_*$. The spin parameter is set to $j=0.7$.}
\label{pic_accumulation}
\end{figure}
We compute the fundamental gravitational-wave (GW) ringdown for a Kerr BH\footnote{The extra factor of $\omega^{-2}$ comes from the second time derivative of $\Psi_4$ to relate it to the strain amplitude $\Psi_4 \sim \ddot{h}$.
Although we could include the spheroidal harmonics ${}_{-2} S_{\ell m} (\omega)$ in the definition of $h_{{\rm G}, \ell m}$, ${}_{-2} S_{\ell m} (\omega_{\ell m n})$ is less sensitive to $n$ for various spin parameters \cite{Oshita:2021iyn}.}:
\begin{equation}
h_{{\rm G},\ell m} (u) = \int_{-\infty}^{\infty} d \omega e^{-i\omega u} \frac{1}{\omega^2} \frac{A^{\rm (T)}_{{\rm out},\ell m}}{2i \omega A^{\rm (T)}_{{\rm in},\ell m}}\,.
\end{equation}
The waveform $h_{{\rm G},\ell m}$ can be expanded with QNMs weighted with the excitation factors $E_{\ell mn}$ at $u > t_*$ as
\begin{equation}
\begin{split}
h_{{\rm E},\ell m} (u) &= \sum_{n_{\rm P}=0}^{N_{\rm P}} E_{\ell mn_{\rm P}} e^{-i\omega_{\ell mn_{\rm P}} (u-t_*)}\\
&+\sum_{n_{\rm R}=0}^{N_{\rm R}} E_{\ell mn_{\rm R}} e^{-i\omega_{\ell mn_{\rm R}} (u-t_*)},
\end{split}
\label{he_timedomain}
\end{equation}
and
\begin{equation}
h_{{\rm G},\ell m} = h_{{\rm E},\ell m} \, \, \text{for} \, \, u > t_*\,,
\label{match}
\end{equation}
when $t_*$ is properly fixed and the number of overtones for prograde/retrograde modes $N_{\rm P}$ and $N_{\rm R}$ is large enough.
In Fig.~\ref{pic_time_shift}, we confirm this explicitly with the mismatch threshold ${\cal M} < {\cal O} (10^{-3})$, where ${\cal M}$ is the mismatch between $h_{{\rm G},\ell m}$ and $h_{{\rm E},\ell m}$:
\begin{equation}
{\cal M} \equiv 1- \left| \frac{\braket{h_{{\rm G},\ell m} | h_{{\rm E},\ell m}}}{\sqrt{\braket{h_{{\rm G},\ell m}|h_{{\rm G},\ell m}} \braket{h_{{\rm E},\ell m}|h_{{\rm E},\ell m}}}} \right|,
\end{equation}
\\
and the braket $\braket{f_1(t)|f_2(t)}$ is the inner product of two functions $f_1(t)$ and $f_2(t)$:
\begin{equation}
\braket{f_1(t)|f_2(t)} \equiv \int_{t_{\rm ini}}^{t_{\rm end}} f_1^{\ast} (t) f_2 (t) dt\,.
\end{equation}

\begin{figure}[ht!]
\centering
\includegraphics[width=1\linewidth]{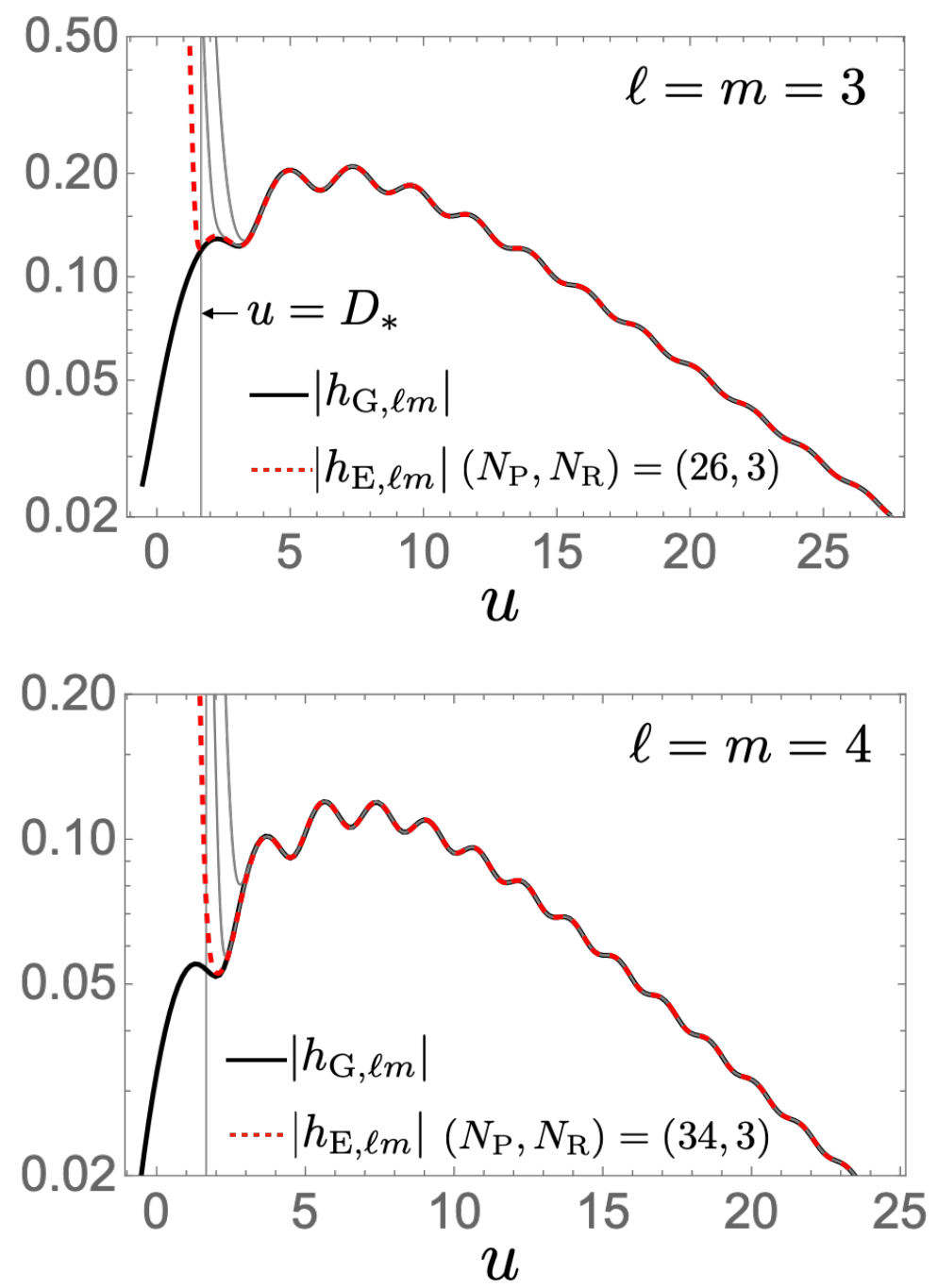}
\caption{Comparison between $h_{{\rm G},22}(u)$ (black) and $h_{{\rm E},22}(u)$ for $\ell=m=3$ (upper) and for $\ell=m=4$ (lower). Gray lines indicate $h_{{\rm E},\ell m}$ with smaller numbers of $N_{\rm P}$ and $N_{\rm R}$: $(N_{\rm P}, N_{\rm R}) = (15,3)$ and $(19,3)$ in the upper panel and $(N_{\rm P}, N_{\rm R}) = (20,3)$ and $(24,3)$ in the lower panel. The vertical gray lines indicate $u=D_*$ as a reference. The spin parameter is set to $j=0.9$.}
\label{pic_lm3344}
\end{figure}
We compute the excitation factors based on Ref. \cite{Oshita:2021iyn} without explicitly introducing the tortoise coordinate and construct $h_{{\rm E},22}$ with $(N_{\rm P},N_{\rm R}) = (20,5)$ and $t_*=0$ (blue solid in Fig.~\ref{pic_time_shift}).
It turns out that the fundamental ringdown $h_{{\rm G},22}$ computed with the conventional tortoise coordinate $r_*$ (black thick line in Fig.~\ref{pic_time_shift}) matches with $h_{{\rm E},22}$ shifted by $t_* = D_*$ (red dashed line in Fig.~\ref{pic_time_shift}), where
\be
D_* = -2 \log(r_+-r_-)\,.\label{D_ana}
\ee
The excitation factors of the Kerr BH in Ref. \cite{Oshita:2021iyn} were obtained by taking the flat limit for the excitation factors of a Kerr de Sitter BH. In this scheme, homogeneous solutions to the Teukolsky equation near the BH horizon ($r=r_+$) and de Sitter one ($r=r_c$) are given by the general Heun function. The imposed boundary condition at the BH horizon in Ref. \cite{Oshita:2021iyn} can be rewritten in the form of $\Delta^2 e^{-ik_{\rm H} \bar{r}_*}$ with another tortoise coordinate $\bar{r}_*$
\begin{align}
\begin{split}
\bar{r}_* &= r + \frac{r_+}{r_+ -r_-} \log \left( \frac{r-r_+}{r_+ - r_-} \right)\\
&-\frac{r_-}{r_+ -r_-} \log \left( \frac{r-r_-}{r_+ - r_-} \right) = r_* + D_*/2\,.
\end{split}
\label{another_tortoise}
\end{align}
The difference between the conventional $r_*$ and $\bar{r}_*$ up to a constant $D_*/2$ leads to the time shift $t_* = D_*$ in $X_{\rm G}$, provided that the mode functions at $r_+ \ll r \ll r_c$ are also defined by $\bar{r}_*$.\footnote{The function $D_* (a)$ diverges in the extremal limit. Therefore, $\bar{r}_*$ defined in (\ref{another_tortoise}) is a more reasonable choice of the tortoise coordinate for which the starting time of the fundamental ringdown is $u=0$.\label{footnote_d_star}}

Figure~\ref{pic_starttime} shows the comparison of $h_{{\rm E},22}$ and $h_{{\rm G},22}$ for several spin parameters. We found that the time shift $t_*$ indeed depends on the spin parameter and is consistent with $t_* = D_*$.
It means that the Green's function or the fundamental ringdown computed in the conventional tortoise coordinate (\ref{conventional_tortoise}) leads to the spin-dependent time shift of ringdown by $t_* = D_* (a)$. We estimate the best-fit value of $t_*$ at which the mismatch between $h_{\rm G}$ and $h_{\rm E}$ takes the least value with ${\cal M} <{\cal O} (10^{-3})$. We find excellent agreement between best-fit $t_*$ and $D_*$ for various spin parameters.

We find that the QNM model $h_{{\rm E},22}$ matches with $h_{{\rm G},22}$ and is convergent (divergent) at $u > t_* = D_*$ (at $u < t_*=D_*$) as shown in Figs.~\ref{pic_time_shift} and \ref{pic_starttime}. This is a direct confirmation of the convergence of the excitation factors and the limitation of the QNM expansion. It also means that the excitation factors in Ref. \cite{Oshita:2021iyn} give the initial amplitude of each QNM in the fundamental ringdown. If we use $\bar{r}_*$ instead of $r_*$, the start time is determined only by the source term as there is no shift caused by the Green's function, i.e., $t_* =0$, in such a case.
Note that the shift of the tortoise coordinate propagates to the source term as well, and the total waveform given by the convolution $h_{{\rm G},\ell m} * h_{{\rm E},\ell m}$ is of course invariant under the transformation of the tortoise coordinate.

The modulation in the fundamental ringdown amplitudes (Figs.~\ref{pic_time_shift} -- \ref{pic_starttime}) is caused by the interference between the prograde and retrograde modes. Also, the beginning of ringdown before the peak can be caused by the destructive interference among multiple overtones, which can be reproduced only when we take into account a significant number of overtones \cite{Oshita:2022pkc}.
These are found by performing the comparison between $h_{{\rm G},\ell m}$ and $h_{{\rm E},\ell m}$ with various numbers of $(N_{\rm P}, N_{\rm R})$ (Fig.~\ref{pic_accumulation}). It is intriguing that one can reconstruct the fundamental ringdown up to around the peak of $|h_{{\rm G},22}|$ for $N_{\rm P}=6$ and $N_{\rm R}=3$. To reconstruct the ringdown even before the peak, one needs to include further higher tones up to around $N_{\rm P} \sim 20$ and $N_{\rm R} \sim 5$ for $j=0.7$. We confirm that the higher the multipole mode is, the more number of QNMs one should include to reconstruct the full ringdown (see also Refs. \cite{Oshita:2021iyn,Oshita:2022pkc}). Figure~\ref{pic_lm3344} shows that for higher multipole modes ($\ell=m=3$ and $4$), one needs more tones to reconstruct the full ringdown, which is consistent with Ref. \cite{Oshita:2022pkc}. Also, the time shift $t_*$ is the same for different multipole modes and overtones, which is because the shift depends on which tortoise coordinate we used to solve the perturbation equation.
Therefore, the $t_*$ depends on the spin parameter $j$ only.

\begin{figure}[t]
\centering
\includegraphics[width=1\linewidth]{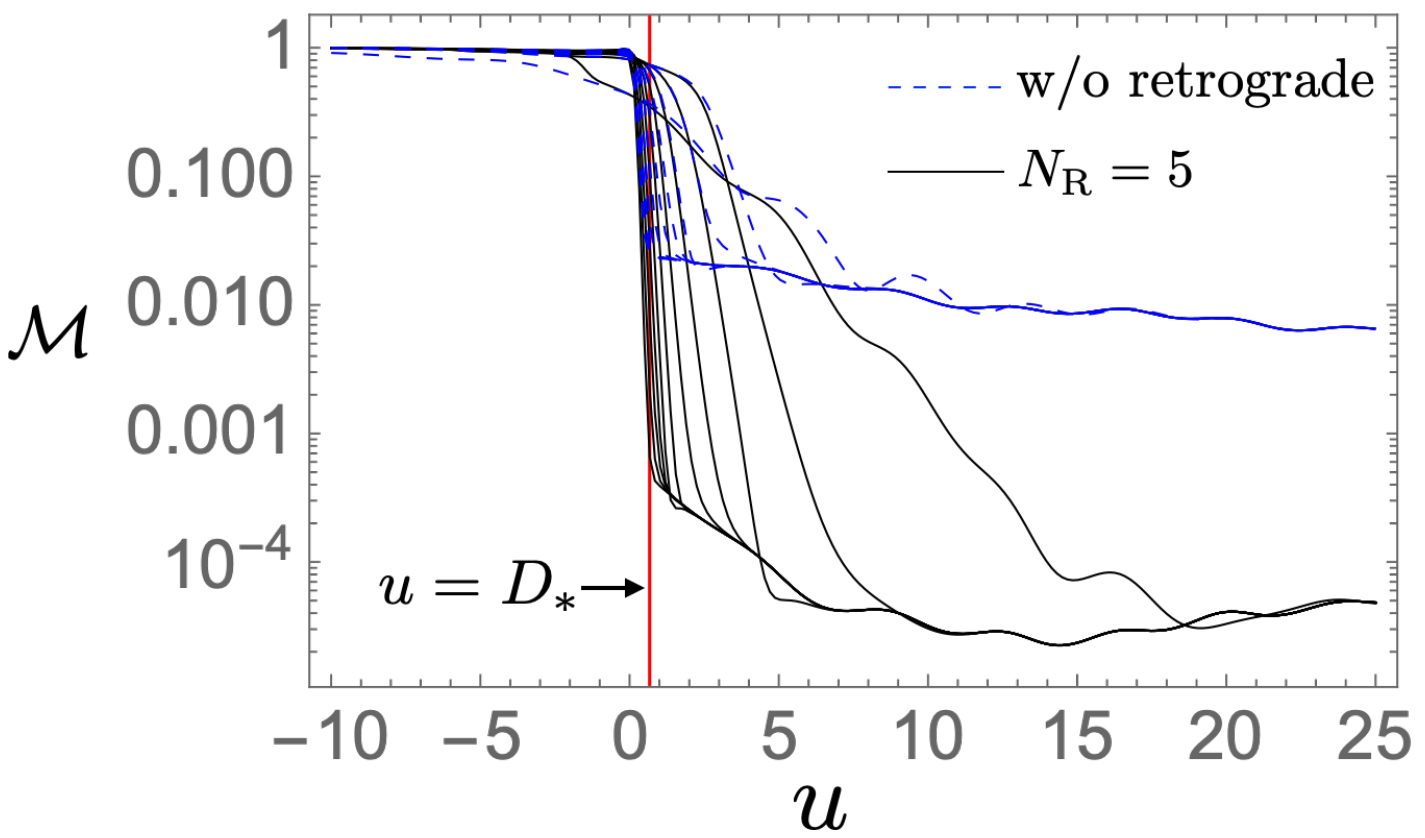}
\caption{Mismatch ${\cal M}$ is shown without retrograde mode (blue dashed) and with $N_{\rm R} =5$ (black solid). The number of prograde modes is $N_{\rm P} = 0, 2, 4, ..., 20$. The red line indicates $u=D_*$. The spin parameter is set to $j=0.7$.}
\label{pic_mismatch}
\end{figure}

\begin{figure}[t]
\centering
\includegraphics[width=1\linewidth]{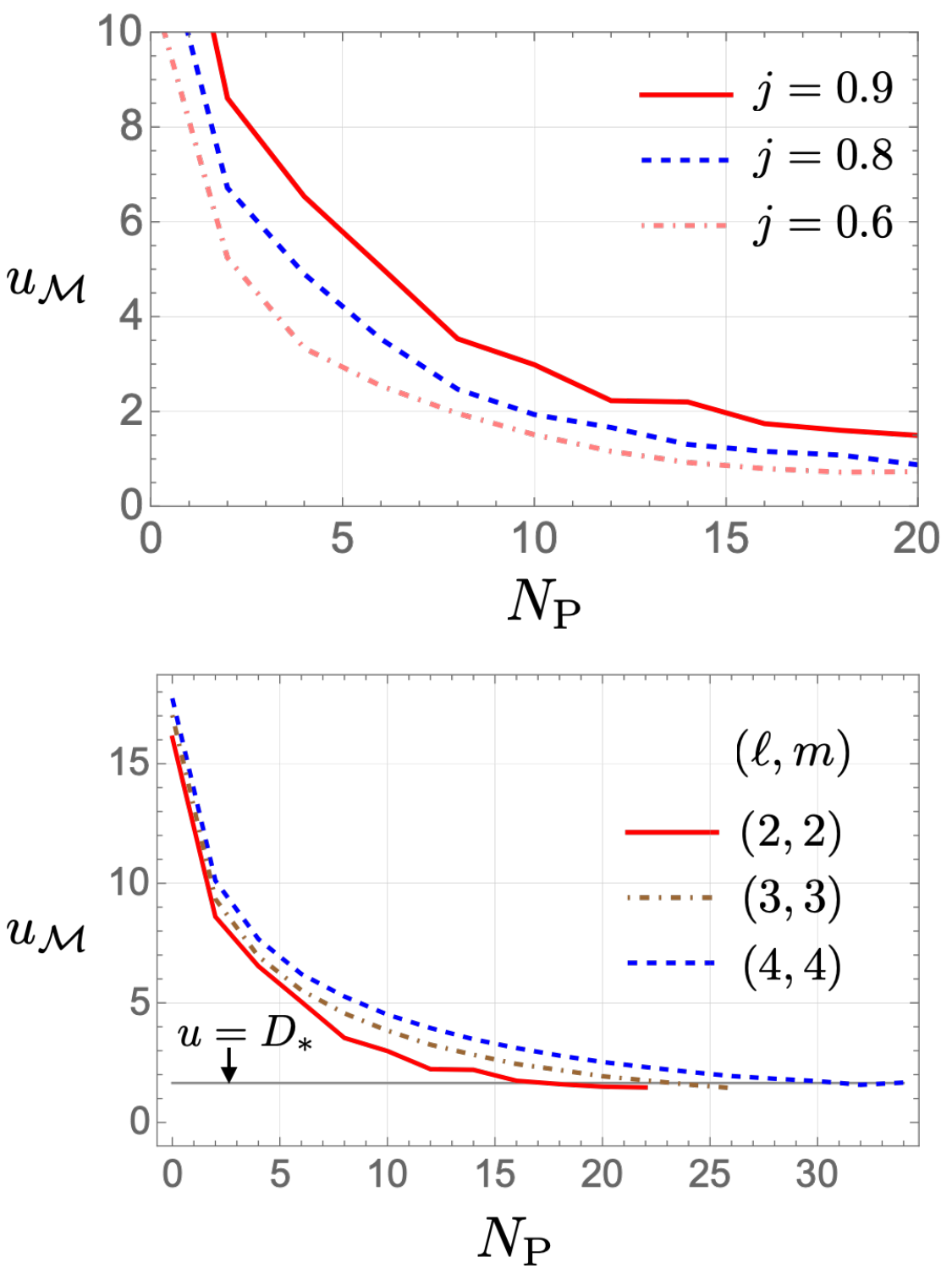}
\caption{Plots of $u_{\cal M}$ at which the mismatch takes ${\cal M} = 5 \times 10^{-4}$. The number of retrograde mode is fixed with $N_{\rm R}=6$, $5$, and $3$ for $j=0.6$, $0.8$, and $0.9$, respectively in the upper panel. In the lower panel, the number of retrograde modes is set to $N_{\rm R} = 3$, and the spin parameter is set to $j=0.9$. The value of $D_*$ is shown as a reference (grey solid).}
\label{pic_mismatchtime}
\end{figure}

To see the convergence of $h_{{\rm E}, \ell m}$ with respect to $N_{\rm P}$ and $N_{\rm R}$, we compute the mismatch ${\cal M}$ of the fundamental ringdown $h_{{\rm G}, \ell m}$ and $h_{{\rm E}, \ell m}$ with $t_*=D_*$.
In Fig.~\ref{pic_mismatch}, we show the mismatch ${\cal M}$ for $-10 \leq t_{\rm ini} \leq 25$ and $t_{\rm end} = 30$. We see that ${\cal M} (u)$ is reduced as the number of $N_{\rm P}$ and $N_{\rm R}$ increases for $u > D_*$.
Figure~\ref{pic_mismatchtime} shows the time $u_{\cal M}$ at which the mismatch is ${\cal M}=5 \times 10^{-4}$. For the quadrupole mode $\ell = m= 2$, the number of prograde overtones $N_{\rm P} \sim 20$ is enough to see the convergence of $u_{\cal M}$. Also, $u_{\cal M}$ is larger for higher spins (see also Fig.~\ref{pic_starttime}). In the lower panel of Fig.~\ref{pic_mismatchtime}, one can read that $u_{\cal M}$ is insensitive to the multipole mode and that the higher the multipole mode is, the more overtones contribute to the fundamental ringdown.

\section{angular mode dependence of the start time of ringdown}
\label{sec_angular_dependence}
An actual ringdown is obtained by the convolution of the fundamental ringdown and a source term. 
In the previous section, we concluded that the time shift caused by the Green's function is the same for different $(\ell,m,n)$ modes. We here discuss the impact of a source term on the starting time of ringdown.
Let us consider the following source term:
\be
I(\omega) = \delta (r_* - x_s) e^{i \phi(\omega)}\,. 
\label{source_second_order}
\ee
We here do not require the reality of this specific source term (\ref{source_second_order}). Our purpose in this section is to discuss and demonstrate how the same starting time of ringdown among $(\ell,m,n)$ is robust when taking into account a non-trivial source term in a simpler setup.
The phase can be expanded with 
\beq
\phi (\omega) =\underbrace{\phi(0)}_{\text{phase shit}} + \underbrace{\phi'(0) \omega}_{\text{time shift $\Delta t$}} + \underbrace{\frac{1}{2} \phi''(0)\omega^2}_{\text{time shift $\Delta t_{\ell m}$}}+ ...
\label{phase}
\eeq
Note that initial data leads to the source term~\eqref{eq:source_wave} which only contains up to linear terms in $\omega$. It is easy to see therefore (comparing $e^{i \phi(\omega)}\sim e^{i \phi(0)}\left(1+i\omega \phi'(0)+...\right)$ with Eq.~\eqref{eq:source_wave}), that for such source
terms with $\omega^2$ and higher are forbidden.
Nevertheless, most problems do have non-initial data-driven source terms. 

We now consider the role of the third term in (\ref{phase}), which leads to the $(\ell,m,n)$-dependent time shift $\Delta t_{\ell mn}$.
We also set $\phi(0) = 0 = \phi'(0)$. From (\ref{source_waveform}), we have
\be
\tilde{X}_{\rm T} (\omega_{\ell mn}) = e^{i\omega_{\ell mn} x_s} e^{i \alpha \omega_{\ell mn}^2}\,,
\ee
where $\alpha \equiv \phi''(0)/2$. 
\begin{figure}[t]
\centering
\includegraphics[width=0.97\linewidth]{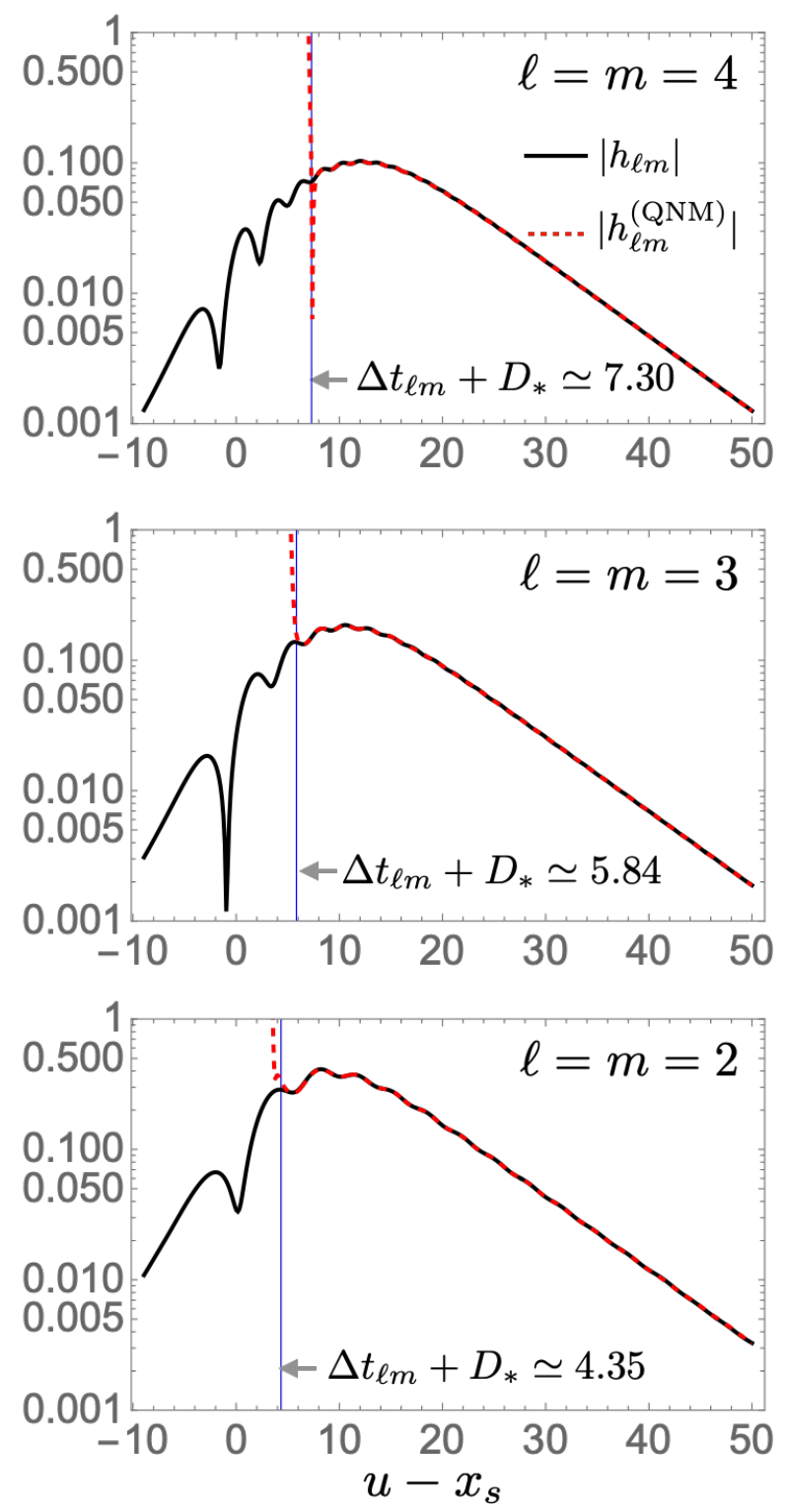}
\caption{Blue vertical lines show the value of $D_* + \Delta t_{\ell m}$ and we here set $\alpha =1$. The spin parameter is set to $j=0.9$.}
\label{pic_lmdependent_starttime}
\end{figure}
The absolute value of the factor $e^{i \alpha \omega_{\ell mn}^2}$ has 
\be
e^{-2 \alpha \text{Re} (\omega_{\ell mn}) \text{Im} (\omega_{\ell mn})}\,,
\label{divergent}
\ee
which diverges at $n \to \infty$ for prograde modes and $\alpha > 0$ or for retrograde modes and $\alpha < 0$.
The ringdown part of the waveform $h_{\ell m}$ is modeled by
\be
h_{\ell m}^{\rm (QNM)} = \sum_{n} E_{\ell mn} e^{-i\omega_{\ell mn} (u -x_s)} e^{i\alpha \omega_{\ell mn}^2}\,.
\ee
All QNMs may be excited, and the factor (\ref{divergent}) is suppressed at
\be
u\geq x_s +D_* + \Delta t_{\ell m} \,,
\label{time_shift_lm}
\ee
with
\be
\Delta t_{\ell m} = 2 \alpha \times \max_n [\text{Re} (\omega_{\ell mn})]\,,
\ee
which depends on the angular mode $(\ell,m)$.
Computing the actual waveform $h_{\ell m} (u)$ with the source term (\ref{source_second_order}) and (\ref{phase}), it is confirmed that the start time depends on $(\ell,m)$ and is consistent with the prediction (\ref{time_shift_lm}) as shown in Fig.~\ref{pic_lmdependent_starttime}.
An actual source term may have a more complicated dependence on $\omega$, which may lead to a different time shift for different $(\ell, m, n)$.
On the other hand, if a source term leads to the phase shift and constant time shift only, e.g., only the first and second terms in (\ref{phase}) are non-zero, the actual excitation time of each QNM is the same for different $(\ell,m,n)$.

\section{Graybody factors and excitation factors}
\label{sec_gray}
\begin{figure*}[t]
\centering
\includegraphics[width=1\linewidth]{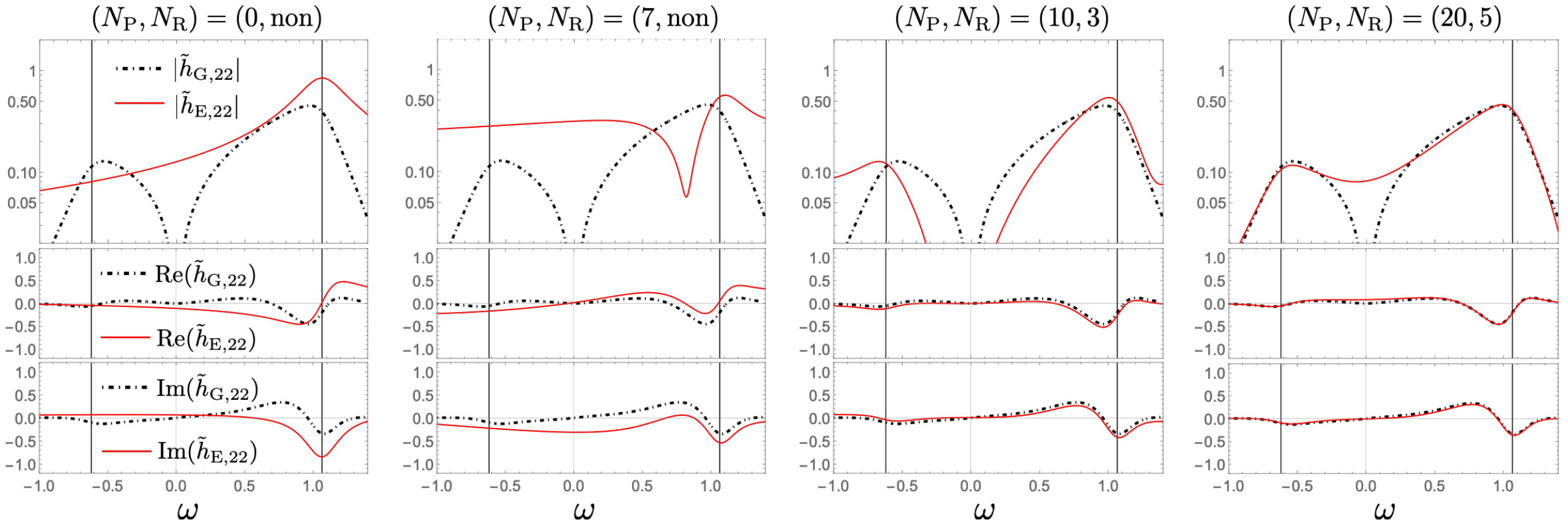}
\caption{Comparison between $\tilde{h}_{{\rm G},\ell m}$ (black dot-dashed) and $\tilde{h}_{{\rm E},\ell m}$ (red solid) with various combination of $(N_{\rm P}, N_{\rm R})$. We set the truncation time $u_0$ in $\tilde{h}_{{\rm E},\ell m}$ as $u_0 = D_*$. The two black vertical lines indicate the QNM frequency of the fundamental prograde and retrograde modes. We set $j=0.7$ and $\ell = m=2$.}
\label{pic_grey_overtones}
\end{figure*}
\begin{figure}[t]
\centering
\includegraphics[width=0.83\linewidth]{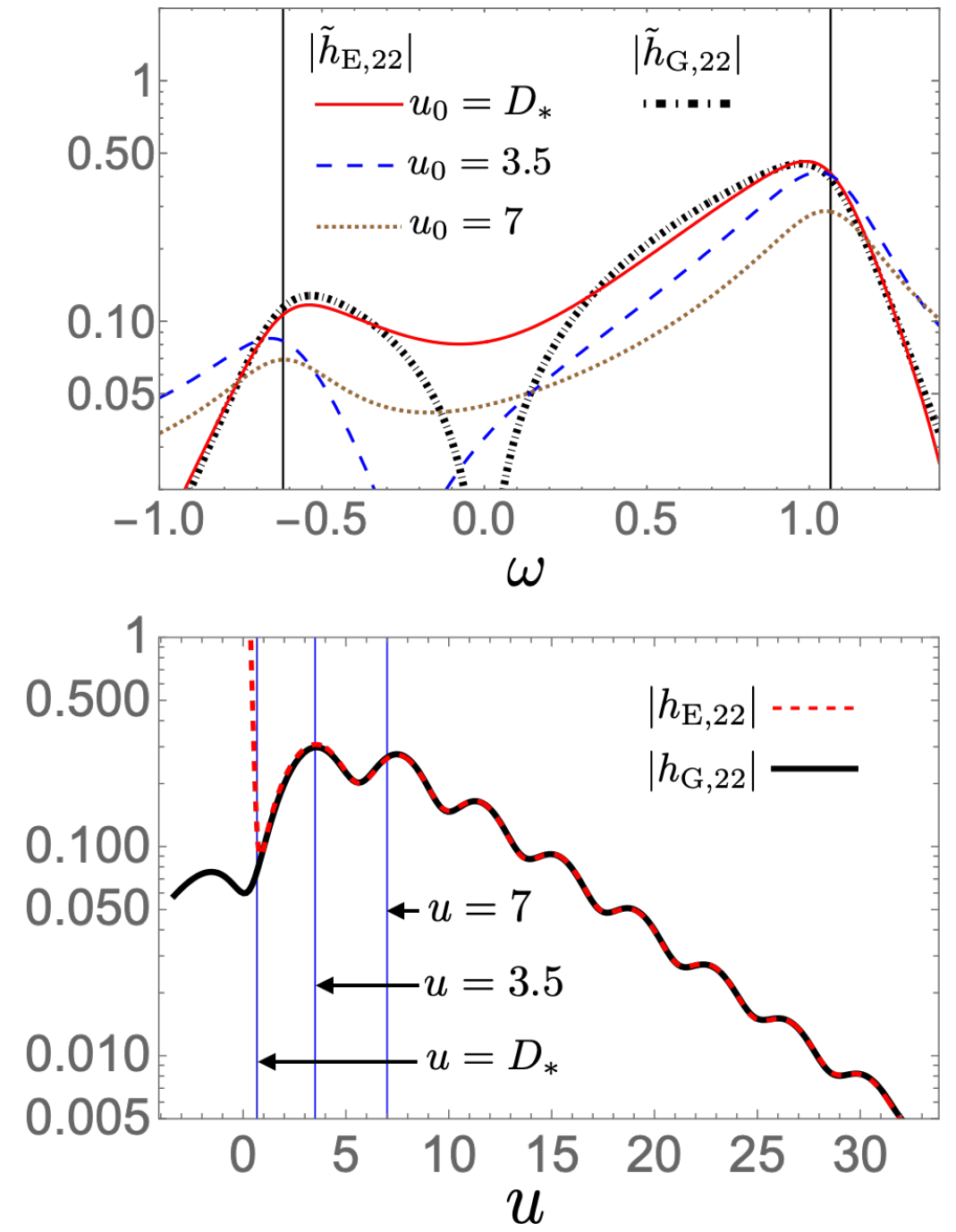}
\caption{{\bf Upper panel:} Comparison between $\tilde{h}_{{\rm G},\ell m}$ (black dot-dashed) and $\tilde{h}_{{\rm E},\ell m}$ with various truncation time $u_0$: $u_0=D_*$ (red solid), $u_0=3.5$ (blue dashed), and $u_0=7$ (brown dotted). We set $j=0.7$ and $\ell = m=2$.
The two black vertical lines indicate the QNM frequency of the fundamental prograde and retrograde modes.
{\bf Lower panel:} Corresponding time-domain waveforms, $h_{{\rm G},\ell m}$ and $h_{{\rm E},\ell m}$ with $t_*=D_*$, are shown. The three corresponding truncation times are shown (blue solid lines) as references.
}
\label{pic_grey_truncation}
\end{figure}
We here study the reconstruction of the spectrum of the fundamental ringdown:
\be
\tilde{h}_{{\rm G},\ell m} = \frac{A_{{\rm out},\ell m}^{\rm (T)}}{2i\omega^3 A_{{\rm in},\ell m}^{\rm (T)}}\,,
\ee
by the QNMs and excitation factors $E_{\ell m n}$.
Figure~\ref{pic_grey_overtones} shows that the high-frequency (ringdown) part of $\tilde{h}_{{\rm G},\ell m}$ can be reproduced with the superposed QNMs weighted with the excitation factors in the frequency domain:
\begin{align}
\begin{split}
\tilde{h}_{{\rm E},\ell m} &=
\int_{u_0}^{\infty} d u h_{{\rm E},\ell m} e^{i\omega u}\\
&= \sum_{n_{\rm P}=0}^{N_{\rm P}} E_{\ell m n_{\rm P}} \frac{i e^{i \omega u_0} e^{-i \omega_{\ell m n_{\rm P}} (u_0-D_*)}}{\omega - \omega_{\ell m n_{\rm P}}}\\
&+\sum_{n_{\rm R}=0}^{N_{\rm R}} E_{\ell m n_{\rm R}} \frac{i e^{i \omega u_0} e^{-i \omega_{\ell m n_{\rm R}} (u_0-D_*)}}{\omega - \omega_{\ell m n_{\rm R}}}\,,
\end{split}
\end{align}
where we set $t_* = D_*$ in (\ref{he_timedomain}) and $u_0$ is the truncation time.
$\tilde{h}_{{\rm E},\ell m}$ truncated at the start time of ringdown, $u_0 = D_*$, exhibits an exponential decay of the spectrum at higher frequencies. The exponential decay is relevant to the BH greybody factor, which has recently been studied as a footprint of the greybody factors in ringdown amplitudes \cite{Oshita:2022pkc,Oshita:2023cjz, Okabayashi:2024qbz}.\footnote{See also Refs. \cite{Rosato:2024arw,Oshita:2024fzf,Konoplya:2024lir}, for the stability of greybody factors against a small correction in the potential barrier.}
One can see that the exponential decay relevant to the feature of greybody factors can be reproduced only when a significant number of QNMs are included in the spectrum.\footnote{See also Ref. \cite{Konoplya:2024lir} for the discussion of the correspondence between the greybody factors and QNMs in the eikonal limit.}
As the prompt response is also included in $\tilde{h}_{{\rm G},\ell m}$, the superposed QNM spectrum $\tilde{h}_{{\rm E},\ell m}$ does not match with $\tilde{h}_{{\rm G},\ell m}$ at lower frequencies.

Also, the reconstruction of the ringdown spectrum, including the exponential decay at high frequencies relevant to the greybody factor, is possible only when we properly set the truncation time $u_0$ at the earliest time of QNM excitation, i.e., $u_0=D_*$, as is shown in Fig.~\ref{pic_grey_truncation} where the superposed QNM spectrum $\tilde{h}_{{\rm E},\ell m}$ is shown for various values of $u_0$.
When we truncate the QNM model $h_{{\rm E},\ell m}$ at later times, e.g. at around the peak amplitude $u_0=3.5$ or even later time $u_0=7$, the exponential decay at higher frequencies in $\tilde{h}_{{\rm G},\ell m}$ cannot be reproduced (Fig. \ref{pic_grey_truncation}).

\section{Conclusion}
\label{sec_conclusion}
We have addressed the time-shift and ringdown starting time problem, by reconstructing the time-domain waveform of ringdown with the excitation factors and quasinormal modes (QNMs), with a specific delta-function source term. We have shown that a significant number of QNMs, including not only prograde modes but also retrograde modes, are necessary to precisely reconstruct the whole ringdown waveform with the mismatch threshold ${\cal M} < {\cal O}(10^{-3})$ (see e.g. Fig.~\ref{pic_accumulation}). For example, for the spin parameter of $j=0.9$ and $\ell =m=2$, we have found that it is enough to include the prograde and retrograde modes up to around the 20th and the third tones, respectively. For higher angular modes, $\ell = m =3$ and $\ell = m =4$, we have found that more overtones of the prograde modes should be included around up to the 26th tone and the 34th tone, respectively (see Figs.~\ref{pic_lm3344} and \ref{pic_mismatchtime}).\footnote{It is necessary to include higher overtones up to $n \sim N$ with $|E_{\ell m 0}| \gtrsim |E_{\ell m N}|$ to see the convergence. The result of excitation factors in Ref. \cite{Oshita:2021iyn} implies that more overtones should be included to reconstruct the fundamental ringdown for lower spins and for near-extremal spins.}
The convergence of the reconstructed ringdown waveform was confirmed by evaluating the mismatch.
We have found that the time shift caused by the Green's function is the same for different $(\ell,m,n)$ modes. On the other hand, depending on the source term we consider, the actual excitation time of each QNM may be distinct even for different angular modes and overtone numbers. 
As a simpler example, we have considered a source term that has a phase factor $e^{i \phi(\omega)}$ and $\phi(\omega)$ has the non-zero higher derivative with respect to $\omega$. Not only the non-trivial phase but also other non-trivial functions, e.g., Gaussian source terms, may also lead to the different excitation times of QNM for different $(\ell,m,n)$ modes. 
We have reconstructed the ringdown waveform with the source term and the excitation factors and have demonstrated that the distinct time shifts are indeed observed for different angular modes (Fig. \ref{pic_lmdependent_starttime}).
If this is not the case, i.e., most of the actual source terms lead to a phase shift and constant time shift only, the actual excitation time is also the same for different $(\ell,m,n)$ modes.
Finally, we have explicitly shown that the exponential decay of the ringdown spectral amplitudes at higher frequencies relevant to the greybody factor can also be reproduced with the superposed QNMs only when a significant number of QNMs are included and the time domain data includes the earliest data of QNM excitation.
To our knowledge, this is the first precise reconstruction of the whole ringdown waveform both in the time and frequency domains, to reveal the start time of ringdown for various values of the spin parameter and angular modes, and to explicitly show that the start time can be different for different angular modes.

Our findings imply that QNMs are not necessarily excited simultaneously when we consider non-trivial complicated source terms, e.g., the origin of the source is highly nonlinear like binary black hole mergers, and the QNM fitting analysis can be performed at different fitting times for different multipole modes. Also, the necessity of including higher prograde overtones and retrograde modes would imply that to reconstruct the ringdown waveform precisely up to its earliest time, one has to include a significant number of QNMs. However, it may lead to the overfitting issue~\cite{Baibhav:2023clw}.
Recently, it has been proposed that the greybody factors would be useful to model the ringdown spectral amplitude for both the source of a plunging particle \cite{Oshita:2023cjz} and the merger of binary BHs \cite{Okabayashi:2024qbz}. Given that the greybody factors are indeed imprinted in ringdown as was argued in Refs. \cite{Oshita:2023cjz,Okabayashi:2024qbz}, our findings imply that a significant number of QNMs are indeed excited at the beginning of ringdown, although it may be challenging to extract each QNM frequency one by one due to the overfitting issue.

\vskip 1cm
\acknowledgments
We thank Emanuele Berti for his comments and for carefully reading the first version of our manuscript.
N.O.~is supported by Japan Society for the Promotion of Science (JSPS) KAKENHI Grant No.~JP23K13111 and by the Hakubi project at Kyoto University.
V.C. acknowledges support by VILLUM Foundation (grant no.\ VIL37766) and the DNRF Chair program (grant no.\ DNRF162) by the Danish National Research Foundation.
V.C.\ acknowledges financial support provided under the European Union’s H2020 ERC Advanced Grant “Black holes: gravitational engines of discovery” grant agreement no.\ Gravitas–101052587. 
Views and opinions expressed are however those of the author only and do not necessarily reflect those of the European Union or the European Research Council. Neither the European Union nor the granting authority can be held responsible for them.
This project has received funding from the European Union's Horizon 2020 research and innovation programme under the Marie Sk{\l}odowska-Curie grant agreement No 101007855 and No 101131233.


\begin{thebibliography}{21}%
\makeatletter
\providecommand \@ifxundefined [1]{%
 \@ifx{#1\undefined}
}%
\providecommand \@ifnum [1]{%
 \ifnum #1\expandafter \@firstoftwo
 \else \expandafter \@secondoftwo
 \fi
}%
\providecommand \@ifx [1]{%
 \ifx #1\expandafter \@firstoftwo
 \else \expandafter \@secondoftwo
 \fi
}%
\providecommand \natexlab [1]{#1}%
\providecommand \enquote  [1]{``#1''}%
\providecommand \bibnamefont  [1]{#1}%
\providecommand \bibfnamefont [1]{#1}%
\providecommand \citenamefont [1]{#1}%
\providecommand \href@noop [0]{\@secondoftwo}%
\providecommand \href [0]{\begingroup \@sanitize@url \@href}%
\providecommand \@href[1]{\@@startlink{#1}\@@href}%
\providecommand \@@href[1]{\endgroup#1\@@endlink}%
\providecommand \@sanitize@url [0]{\catcode `\\12\catcode `\$12\catcode
  `\&12\catcode `\#12\catcode `\^12\catcode `\_12\catcode `\%12\relax}%
\providecommand \@@startlink[1]{}%
\providecommand \@@endlink[0]{}%
\providecommand \url  [0]{\begingroup\@sanitize@url \@url }%
\providecommand \@url [1]{\endgroup\@href {#1}{\urlprefix }}%
\providecommand \urlprefix  [0]{URL }%
\providecommand \Eprint [0]{\href }%
\providecommand \doibase [0]{http://dx.doi.org/}%
\providecommand \selectlanguage [0]{\@gobble}%
\providecommand \bibinfo  [0]{\@secondoftwo}%
\providecommand \bibfield  [0]{\@secondoftwo}%
\providecommand \translation [1]{[#1]}%
\providecommand \BibitemOpen [0]{}%
\providecommand \bibitemStop [0]{}%
\providecommand \bibitemNoStop [0]{.\EOS\space}%
\providecommand \EOS [0]{\spacefactor3000\relax}%
\providecommand \BibitemShut  [1]{\csname bibitem#1\endcsname}%
\let\auto@bib@innerbib\@empty
\bibitem [{\citenamefont {Vishveshwara}(1970)}]{Vishveshwara:1970zz}%
  \BibitemOpen
  \bibfield  {author} {\bibinfo {author} {\bibfnamefont {C.~V.}\ \bibnamefont
  {Vishveshwara}},\ }\href {\doibase 10.1038/227936a0} {\bibfield  {journal}
  {\bibinfo  {journal} {Nature}\ }\textbf {\bibinfo {volume} {227}},\ \bibinfo
  {pages} {936} (\bibinfo {year} {1970})}\BibitemShut {NoStop}%
\bibitem [{\citenamefont {Chandrasekhar}\ and\ \citenamefont
  {Detweiler}(1975)}]{Chandrasekhar:1975zza}%
  \BibitemOpen
  \bibfield  {author} {\bibinfo {author} {\bibfnamefont {S.}~\bibnamefont
  {Chandrasekhar}}\ and\ \bibinfo {author} {\bibfnamefont {S.~L.}\ \bibnamefont
  {Detweiler}},\ }\href {\doibase 10.1098/rspa.1975.0112} {\bibfield  {journal}
  {\bibinfo  {journal} {Proc. Roy. Soc. Lond. A}\ }\textbf {\bibinfo {volume}
  {344}},\ \bibinfo {pages} {441} (\bibinfo {year} {1975})}\BibitemShut
  {NoStop}%
\bibitem [{\citenamefont {Berti}\ \emph {et~al.}(2009)\citenamefont {Berti},
  \citenamefont {Cardoso},\ and\ \citenamefont {Starinets}}]{Berti:2009kk}%
  \BibitemOpen
  \bibfield  {author} {\bibinfo {author} {\bibfnamefont {E.}~\bibnamefont
  {Berti}}, \bibinfo {author} {\bibfnamefont {V.}~\bibnamefont {Cardoso}}, \
  and\ \bibinfo {author} {\bibfnamefont {A.~O.}\ \bibnamefont {Starinets}},\
  }\href {\doibase 10.1088/0264-9381/26/16/163001} {\bibfield  {journal}
  {\bibinfo  {journal} {Class. Quant. Grav.}\ }\textbf {\bibinfo {volume}
  {26}},\ \bibinfo {pages} {163001} (\bibinfo {year} {2009})},\ \Eprint
  {http://arxiv.org/abs/0905.2975} {arXiv:0905.2975 [gr-qc]} \BibitemShut
  {NoStop}%
\bibitem [{\citenamefont {Baibhav}\ \emph {et~al.}(2023)\citenamefont
  {Baibhav}, \citenamefont {Cheung}, \citenamefont {Berti}, \citenamefont
  {Cardoso}, \citenamefont {Carullo}, \citenamefont {Cotesta}, \citenamefont
  {Del~Pozzo},\ and\ \citenamefont {Duque}}]{Baibhav:2023clw}%
  \BibitemOpen
  \bibfield  {author} {\bibinfo {author} {\bibfnamefont {V.}~\bibnamefont
  {Baibhav}}, \bibinfo {author} {\bibfnamefont {M.~H.-Y.}\ \bibnamefont
  {Cheung}}, \bibinfo {author} {\bibfnamefont {E.}~\bibnamefont {Berti}},
  \bibinfo {author} {\bibfnamefont {V.}~\bibnamefont {Cardoso}}, \bibinfo
  {author} {\bibfnamefont {G.}~\bibnamefont {Carullo}}, \bibinfo {author}
  {\bibfnamefont {R.}~\bibnamefont {Cotesta}}, \bibinfo {author} {\bibfnamefont
  {W.}~\bibnamefont {Del~Pozzo}}, \ and\ \bibinfo {author} {\bibfnamefont
  {F.}~\bibnamefont {Duque}},\ }\href {\doibase 10.1103/PhysRevD.108.104020}
  {\bibfield  {journal} {\bibinfo  {journal} {Phys. Rev. D}\ }\textbf {\bibinfo
  {volume} {108}},\ \bibinfo {pages} {104020} (\bibinfo {year} {2023})},\
  \Eprint {http://arxiv.org/abs/2302.03050} {arXiv:2302.03050 [gr-qc]}
  \BibitemShut {NoStop}%
\bibitem [{\citenamefont {Andersson}(1997)}]{Andersson:1996cm}%
  \BibitemOpen
  \bibfield  {author} {\bibinfo {author} {\bibfnamefont {N.}~\bibnamefont
  {Andersson}},\ }\href {\doibase 10.1103/PhysRevD.55.468} {\bibfield
  {journal} {\bibinfo  {journal} {Phys. Rev. D}\ }\textbf {\bibinfo {volume}
  {55}},\ \bibinfo {pages} {468} (\bibinfo {year} {1997})},\ \Eprint
  {http://arxiv.org/abs/gr-qc/9607064} {arXiv:gr-qc/9607064} \BibitemShut
  {NoStop}%
\bibitem [{\citenamefont {Nollert}\ and\ \citenamefont
  {Price}(1999)}]{Nollert:1998ys}%
  \BibitemOpen
  \bibfield  {author} {\bibinfo {author} {\bibfnamefont {H.-P.}\ \bibnamefont
  {Nollert}}\ and\ \bibinfo {author} {\bibfnamefont {R.~H.}\ \bibnamefont
  {Price}},\ }\href {\doibase 10.1063/1.532698} {\bibfield  {journal} {\bibinfo
   {journal} {J. Math. Phys.}\ }\textbf {\bibinfo {volume} {40}},\ \bibinfo
  {pages} {980} (\bibinfo {year} {1999})},\ \Eprint
  {http://arxiv.org/abs/gr-qc/9810074} {arXiv:gr-qc/9810074} \BibitemShut
  {NoStop}%
\bibitem [{\citenamefont {Sun}\ and\ \citenamefont {Price}(1988)}]{Sun:1988tz}%
  \BibitemOpen
  \bibfield  {author} {\bibinfo {author} {\bibfnamefont {Y.}~\bibnamefont
  {Sun}}\ and\ \bibinfo {author} {\bibfnamefont {R.~H.}\ \bibnamefont
  {Price}},\ }\href {\doibase 10.1103/PhysRevD.38.1040} {\bibfield  {journal}
  {\bibinfo  {journal} {Phys. Rev. D}\ }\textbf {\bibinfo {volume} {38}},\
  \bibinfo {pages} {1040} (\bibinfo {year} {1988})}\BibitemShut {NoStop}%
\bibitem [{\citenamefont {Berti}\ and\ \citenamefont
  {Cardoso}(2006)}]{Berti:2006wq}%
  \BibitemOpen
  \bibfield  {author} {\bibinfo {author} {\bibfnamefont {E.}~\bibnamefont
  {Berti}}\ and\ \bibinfo {author} {\bibfnamefont {V.}~\bibnamefont
  {Cardoso}},\ }\href {\doibase 10.1103/PhysRevD.74.104020} {\bibfield
  {journal} {\bibinfo  {journal} {Phys. Rev. D}\ }\textbf {\bibinfo {volume}
  {74}},\ \bibinfo {pages} {104020} (\bibinfo {year} {2006})},\ \Eprint
  {http://arxiv.org/abs/gr-qc/0605118} {arXiv:gr-qc/0605118} \BibitemShut
  {NoStop}%
\bibitem [{\citenamefont {Leaver}(1986{\natexlab{a}})}]{Leaver:1986gd}%
  \BibitemOpen
  \bibfield  {author} {\bibinfo {author} {\bibfnamefont {E.~W.}\ \bibnamefont
  {Leaver}},\ }\href {\doibase 10.1103/PhysRevD.34.384} {\bibfield  {journal}
  {\bibinfo  {journal} {Phys. Rev. D}\ }\textbf {\bibinfo {volume} {34}},\
  \bibinfo {pages} {384} (\bibinfo {year} {1986}{\natexlab{a}})}\BibitemShut
  {NoStop}%
\bibitem [{\citenamefont {Leaver}(1985)}]{Leaver:1985ax}%
  \BibitemOpen
  \bibfield  {author} {\bibinfo {author} {\bibfnamefont {E.~W.}\ \bibnamefont
  {Leaver}},\ }\href {\doibase 10.1098/rspa.1985.0119} {\bibfield  {journal}
  {\bibinfo  {journal} {Proc. Roy. Soc. Lond. A}\ }\textbf {\bibinfo {volume}
  {402}},\ \bibinfo {pages} {285} (\bibinfo {year} {1985})}\BibitemShut
  {NoStop}%
\bibitem [{\citenamefont {Leaver}(1986{\natexlab{b}})}]{Leaver:1986vnb}%
  \BibitemOpen
  \bibfield  {author} {\bibinfo {author} {\bibfnamefont {E.~W.}\ \bibnamefont
  {Leaver}},\ }\href {\doibase 10.1063/1.527130} {\bibfield  {journal}
  {\bibinfo  {journal} {J. Math. Phys.}\ }\textbf {\bibinfo {volume} {27}},\
  \bibinfo {pages} {1238} (\bibinfo {year} {1986}{\natexlab{b}})}\BibitemShut
  {NoStop}%
\bibitem [{\citenamefont {Andersson}(1995)}]{Andersson:1995zk}%
  \BibitemOpen
  \bibfield  {author} {\bibinfo {author} {\bibfnamefont {N.}~\bibnamefont
  {Andersson}},\ }\href {\doibase 10.1103/PhysRevD.51.353} {\bibfield
  {journal} {\bibinfo  {journal} {Phys. Rev. D}\ }\textbf {\bibinfo {volume}
  {51}},\ \bibinfo {pages} {353} (\bibinfo {year} {1995})}\BibitemShut
  {NoStop}%
\bibitem [{\citenamefont {Zhang}\ \emph {et~al.}(2013)\citenamefont {Zhang},
  \citenamefont {Berti},\ and\ \citenamefont {Cardoso}}]{Zhang:2013ksa}%
  \BibitemOpen
  \bibfield  {author} {\bibinfo {author} {\bibfnamefont {Z.}~\bibnamefont
  {Zhang}}, \bibinfo {author} {\bibfnamefont {E.}~\bibnamefont {Berti}}, \ and\
  \bibinfo {author} {\bibfnamefont {V.}~\bibnamefont {Cardoso}},\ }\href
  {\doibase 10.1103/PhysRevD.88.044018} {\bibfield  {journal} {\bibinfo
  {journal} {Phys. Rev. D}\ }\textbf {\bibinfo {volume} {88}},\ \bibinfo
  {pages} {044018} (\bibinfo {year} {2013})},\ \Eprint
  {http://arxiv.org/abs/1305.4306} {arXiv:1305.4306 [gr-qc]} \BibitemShut
  {NoStop}%
\bibitem [{\citenamefont {Oshita}(2021)}]{Oshita:2021iyn}%
  \BibitemOpen
  \bibfield  {author} {\bibinfo {author} {\bibfnamefont {N.}~\bibnamefont
  {Oshita}},\ }\href {\doibase 10.1103/PhysRevD.104.124032} {\bibfield
  {journal} {\bibinfo  {journal} {Phys. Rev. D}\ }\textbf {\bibinfo {volume}
  {104}},\ \bibinfo {pages} {124032} (\bibinfo {year} {2021})},\ \Eprint
  {http://arxiv.org/abs/2109.09757} {arXiv:2109.09757 [gr-qc]} \BibitemShut
  {NoStop}%
\bibitem [{\citenamefont {Zhu}\ \emph {et~al.}(2024)\citenamefont {Zhu},
  \citenamefont {Ripley}, \citenamefont {C\'ardenas-Avenda\~no},\ and\
  \citenamefont {Pretorius}}]{Zhu:2023mzv}%
  \BibitemOpen
  \bibfield  {author} {\bibinfo {author} {\bibfnamefont {H.}~\bibnamefont
  {Zhu}}, \bibinfo {author} {\bibfnamefont {J.~L.}\ \bibnamefont {Ripley}},
  \bibinfo {author} {\bibfnamefont {A.}~\bibnamefont {C\'ardenas-Avenda\~no}},
  \ and\ \bibinfo {author} {\bibfnamefont {F.}~\bibnamefont {Pretorius}},\
  }\href {\doibase 10.1103/PhysRevD.109.044010} {\bibfield  {journal} {\bibinfo
   {journal} {Phys. Rev. D}\ }\textbf {\bibinfo {volume} {109}},\ \bibinfo
  {pages} {044010} (\bibinfo {year} {2024})},\ \Eprint
  {http://arxiv.org/abs/2309.13204} {arXiv:2309.13204 [gr-qc]} \BibitemShut
  {NoStop}%
\bibitem [{\citenamefont {Oshita}(2023)}]{Oshita:2022pkc}%
  \BibitemOpen
  \bibfield  {author} {\bibinfo {author} {\bibfnamefont {N.}~\bibnamefont
  {Oshita}},\ }\href {\doibase 10.1088/1475-7516/2023/04/013} {\bibfield
  {journal} {\bibinfo  {journal} {JCAP}\ }\textbf {\bibinfo {volume} {04}},\
  \bibinfo {pages} {013} (\bibinfo {year} {2023})},\ \Eprint
  {http://arxiv.org/abs/2208.02923} {arXiv:2208.02923 [gr-qc]} \BibitemShut
  {NoStop}%
\bibitem [{\citenamefont {Oshita}(2024)}]{Oshita:2023cjz}%
  \BibitemOpen
  \bibfield  {author} {\bibinfo {author} {\bibfnamefont {N.}~\bibnamefont
  {Oshita}},\ }\href {\doibase 10.1103/PhysRevD.109.104028} {\bibfield
  {journal} {\bibinfo  {journal} {Phys. Rev. D}\ }\textbf {\bibinfo {volume}
  {109}},\ \bibinfo {pages} {104028} (\bibinfo {year} {2024})},\ \Eprint
  {http://arxiv.org/abs/2309.05725} {arXiv:2309.05725 [gr-qc]} \BibitemShut
  {NoStop}%
\bibitem [{\citenamefont {Okabayashi}\ and\ \citenamefont
  {Oshita}(2024)}]{Okabayashi:2024qbz}%
  \BibitemOpen
  \bibfield  {author} {\bibinfo {author} {\bibfnamefont {K.}~\bibnamefont
  {Okabayashi}}\ and\ \bibinfo {author} {\bibfnamefont {N.}~\bibnamefont
  {Oshita}},\ }\href@noop {} {\  (\bibinfo {year} {2024})},\ \Eprint
  {http://arxiv.org/abs/2403.17487} {arXiv:2403.17487 [gr-qc]} \BibitemShut
  {NoStop}%
\bibitem [{\citenamefont {Rosato}\ \emph {et~al.}(2024)\citenamefont {Rosato},
  \citenamefont {Destounis},\ and\ \citenamefont {Pani}}]{Rosato:2024arw}%
  \BibitemOpen
  \bibfield  {author} {\bibinfo {author} {\bibfnamefont {R.~F.}\ \bibnamefont
  {Rosato}}, \bibinfo {author} {\bibfnamefont {K.}~\bibnamefont {Destounis}}, \
  and\ \bibinfo {author} {\bibfnamefont {P.}~\bibnamefont {Pani}},\ }\href@noop
  {} {\  (\bibinfo {year} {2024})},\ \Eprint {http://arxiv.org/abs/2406.01692}
  {arXiv:2406.01692 [gr-qc]} \BibitemShut {NoStop}%
\bibitem [{\citenamefont {Oshita}\ \emph {et~al.}(2024)\citenamefont {Oshita},
  \citenamefont {Takahashi},\ and\ \citenamefont {Mukohyama}}]{Oshita:2024fzf}%
  \BibitemOpen
  \bibfield  {author} {\bibinfo {author} {\bibfnamefont {N.}~\bibnamefont
  {Oshita}}, \bibinfo {author} {\bibfnamefont {K.}~\bibnamefont {Takahashi}}, \
  and\ \bibinfo {author} {\bibfnamefont {S.}~\bibnamefont {Mukohyama}},\
  }\href@noop {} {\  (\bibinfo {year} {2024})},\ \Eprint
  {http://arxiv.org/abs/2406.04525} {arXiv:2406.04525 [gr-qc]} \BibitemShut
  {NoStop}%
\bibitem [{\citenamefont {Konoplya}\ and\ \citenamefont
  {Zhidenko}(2024)}]{Konoplya:2024lir}%
  \BibitemOpen
  \bibfield  {author} {\bibinfo {author} {\bibfnamefont {R.~A.}\ \bibnamefont
  {Konoplya}}\ and\ \bibinfo {author} {\bibfnamefont {A.}~\bibnamefont
  {Zhidenko}},\ }\href@noop {} {\  (\bibinfo {year} {2024})},\ \Eprint
  {http://arxiv.org/abs/2406.11694} {arXiv:2406.11694 [gr-qc]} \BibitemShut
  {NoStop}%
\end{thebibliography}
\end{document}